\documentclass[12pt]{elsart}
\usepackage{latexsym}
\usepackage{amsfonts,amssymb,amstext}
\renewcommand{\d}{\partial}

\newcommand{\C}{{\ensuremath{\mathbb C}}}

\newcommand{\R}{{\ensuremath{\mathbb R}}}

\newcommand{\X}{{\ensuremath{\mathsf X}}}
\newcommand{\Y}{{\ensuremath{\mathsf Y}}}
\newcommand{\Z}{{\ensuremath{\mathsf Z}}}
\newcommand{\B}{{\ensuremath{\mathsf B}}}
\newcommand{\A}{{\ensuremath{\mathsf A}}}
\newcommand{\HH}{{\ensuremath{\mathsf H}}}
\newcommand{\ee}{{\ensuremath{\mathsf e}}}
\newcommand{\bsf}{{\ensuremath{\mathsf b}}}
\newcommand{\asf}{{\ensuremath{\mathsf a}}}
\newcommand{\nn}{{\ensuremath{\mathbf n}}}
\newcommand{\x}{{\ensuremath{\mathbf x}}}
\newcommand{\p}{{\ensuremath{\mathbf p}}}
\newcommand{\PP}{\mbox{\boldmath $\psi$}}
\newcommand{\N}{\mbox{\boldmath $\eta$}}
\newcommand{\balpha}{\mbox{\boldmath $\alpha$}}
\newcommand{\bbalpha}{\mbox{\boldmath $\bar\alpha$}}

\newcommand{\NPB}[3]{{\sl Nucl. Phys.} {\bf B#1} (#2) #3}

\newcommand{\CQG}[3]{{\sl Class. Quantum Grav.} {\bf #1} (#2) #3}

\newcommand{\JDG}[3] {{\sl J. Diff. Geom.} {\bf #1} (#2) #3}

\newcommand{\AIHP}[3] {{\sl Ann. Inst.
Henri Poincar\' e} {\bf #1} (#2) #3}

\begin{document}
\begin{frontmatter}
\title{The reduced covariant phase space 
quantization of the three dimensional Nambu-Goto string}
\author[UAM]{Eduardo Ramos\thanksref{emailed}}

\address[UAM]{Dept. de F\'{\i}sica Te\'orica, C-XI\break
                          Universidad Aut\'onoma de Madrid\break
                         Ciudad Universitaria de Cantoblanco\break
                          28049 Madrid, SPAIN}        

\thanks[emailed]{\tt mailto:ramos@delta.ft.uam.es}

\begin{abstract}
The reduced covariant phase space associated with the
three-dimensional Euclidean Nambu-Goto action can be identified, via
the Enneper-Weierstrass representation of minimal surfaces, with the
space of complex analytic functions plus three translational zero
modes. The symplectic structure induced trough the Enneper-Weierstrass
map can be explicitly computed. Quantization is then straightforward,
yielding as a result a target-space Euclidean-invariant,
positive-definite, two-dimensional quantum field theory.  The physical
states are shown to correspond with particles states of integer spin
and arbitrary mass.
\end{abstract}
\end{frontmatter}

\section{Introduction}

String theory is certainly a fascinating topic. It aims towards a complete
understanding of the principles of physical interaction solely in
terms of geometry. Nevertheless, as could have been easily foreseen,
such an ambitious program is plagued with almost unsurmountable
difficulties.  Although much progress has been achieved since the epic
times of the first dual models of strong interactions, when the first
hints were provided that a string model could be useful in
understanding fundamental physical process, much
still remains to be done. For example, the Nambu-Goto string
model, the simplest and oldest among all of them, is still waiting to
be consistently quantized outside the critical dimension. It is the
purpose of this work to report on some progress in that direction.

In the present work I will restrict myself to the three-dimensional
case. The reasons to do so are manifold. First of all the theory of
immersed surfaces is a subject that has been thoroughly studied since
the pioneering days of Gauss, and therefore a plethora of deep
geometrical results are directly available. This work will be directly
based upon them; but this is certainly not the only reason.  Three
dimensions play in the case of strings the analogue of two dimensions in
the particle case, i.e., the lowest dimension for which dynamics are
not trivial in nature; it is therefore natural to expect it to be the
simpler case. It is by now a futile exercise to try to justify the
important role that two-dimensional quantum field theory has played in
our current understanding of more realistic particle theories. It is
my believe that three dimensional string theory must occupy a similar
relevant place in the quantum theory of extended objects. Nevertheless
it is clear that the importance of three-dimensional string theory
transcends the limits of the so-called ``theory of everything'' and
may find direct application in several, and perhaps more realistic,
physical situations. The statistical mechanics of membranes,
interfaces, and spin systems range among the most important among
them. The transfer matrix formalism allows for a direct application of
the quantum mechanics of three dimensional strings to this statistical
mechanical models, and viceversa. In particular, it has been argued by
Polyakov that the three-dimensional Ising model near criticality
should be described by a three-dimensional string theory.

There have been traditionally two different approaches to the
quantization of strings. The covariant approach, by far the most
popular nowadays, is obtained by implementing the constraints coming
from reparameterization invariance \`a la Dirac, or in more modern
treatments via the construction of its associated BRST charge. It is
well known that consistency of this approach requires the dimension of
space-time to be 26. This fact induced Polyakov \cite{Polyakov} to
consider an alternative, and still covariant, line of attack based on
the coupling of conformal matter to two-dimensional gravity. Although
classically both approaches are easily seen to be equivalent for the
case of $D$ bosons coupled to two-dimensional gravity (where $D$
represents the dimension of the target space), Polyakov's treatment 
permitted, trough a careful study of the Weyl anomaly, to extend the
analysis to the non-critical case.  Unfortunately, the existence of
the infamous $c=1$ barrier has not yet allowed us to study the
physically interesting dimensions, arguably 3 and 4.

The other main line of attack to the problem is the so-called
light-cone gauge quantization. The idea behind it is to completely
reduce the phase space of the string theory and quantized directly its
physical transverse modes. There is an ungoing discussion in the
literature about which is, on general principles, the ``correct'' way
to quantize a dynamical system subject to constraints. There are many
finite-dimensional examples for which the Dirac and the reduced phase
space quantization algorithms yield different, but yet consistent,
quantum theories.  Nevertheless, it seems that physical input
determines as the correct algorithm one or the other, depending on the
particular model, and that there is no particular one a priori
prefered by nature. In the light-cone approach the reduced phase space
is obtained trough the choice of a non Poincar\'e covariant gauge
condition. Roughly speaking, one of the coordinates in the world sheet
is fixed to correspond to the time coordinate of the space-time in
which the string lives. In this approach the magic number 26 pops out
as the only dimension in which one can construct a representation of
the Poincar\'e algebra. The technical reasons behind this coming from
the fact that due to the non-covariant gauge fixing the Poincar\'e
group is nonlinearly realized at the field level, provoking ordering
problems that seem only to be fixed for the critical dimension.

The approach that I will present here is closer in nature to the
light-cone approach, in the sense that the quantization procedure will
go trough a complete reduction of the phase space of the Nambu-Goto
string prior to quantization. Nevertheless, and in contrast to the
light-cone approach, it will be possible to keep explicitly
Poincar\'e, or rather Euclidean, invariance all along the way. The key
ingredient to the construction will be the Enneper-Weierstrass
representation of minimal surfaces. As I will show, following
otherwise completely standard geometrical constructions, the reduced
phase space of the three-dimensional closed Nambu-Goto string can be
locally identified with the space of complex analytic functions plus
three translational zero modes.  In the reduction process the
conformal structure is completely fixed by choosing a geometrical
parameterization of the surface in terms of its Gauss map by
stereographic projection.  As a consequence of this, rotational
invariance is explicitly implemented as a $SU(2)$ subgroup of the
standard linear fractional transformations acting on the Riemann
sphere, and realized linearly on the physical fields.

Of course, quantization requires something more than the
identification of the reduced phase space, an explicit expression for
the induced symplectic structure is mandatory.  Surprisingly enough
this will prove to be a simple task. The required machinery is
provided by the covariant phase space approach to hamiltonian
mechanics together with some basic symplectic geometry. Even though we
will be working with infinite-dimensional manifolds, an adequate
algebraization of the required geometrical constructions will allow us
to carry out the reduction process in a rather standard fashion.

The plan of the paper is as follows. First I will remind the reader of
some basic notions about the geometry of immersed surfaces in $\R^3$
that will be used in the following. Then I will introduce the
Enneper-Weierstrass representation of minimal surfaces, and I will
show how Euclidean invariance is realized within this representation.

Next, I will briefly recall some general results about the
three-dimensional Nambu-Goto string and how they fit in the context of
minimal surface theory. After this, and in order to understand how to
define a symplectic structure in the reduced phase space of the
theory, I will digress on the covariant phase space approach to
classical mechanics, and I will explicitly show how to apply it in
this particular case.

Finally, I will attack the quantization of the model. I will show,
that in contrast to the standard approach, the quantization in this
case yields as a result a target-space Euclidean-invariant,
positive-definite, two-dimensional quantum field theory. The
associated physical states, which are obtained by the repeated action
of creation operators on the vacuum state, can then be identified, via
the Wigner method of induced representations applied to the Euclidean
group, with particle states of integer spin and arbitrary mass.

\section{A very brief course about surface theory in $\R^3$}

The purpose of this introductory section is to present in a simple
manner the most important geometrical constructions to be used in the
sequel, as well as to set up my notations.  For a comprehensive
introduction to this fascinating subject I refer the reader to the
excellent book of M.~Spivak \cite{Spivak}.

Let $\Sigma$ be an oriented two-dimensional connected Riemannian
manifold and $\X: \Sigma\rightarrow\R^3$ an isometric immersion of
$\Sigma$ into $\R^3$.  At any point $p$ of $\Sigma$ a basis for the
tangent plane is provided by $\partial_{\alpha} X^{i}$. The induced
metric, or first fundamental form of the immersion, is then given by
\begin{equation}
g_{\alpha\beta} = \partial_{\alpha}\X\cdot\partial_{\beta}\X.
\end{equation}
It is now possible to obtain a basis for $T\R^3$ at $p$ by adding a
unitary perpendicular vector $\nn$, whose explicit coordinate
expression may be given by
\begin{equation}
n^{i} = {1\over 2\sqrt{g}} \epsilon^{ijk}\epsilon^{\alpha\beta}
\d_{\alpha}X^{j}\d_{\beta}X^{k},
\end{equation}
with $g$ being the determinant of the induced metric.

One may now write down the structural equations of the immersion as
\begin{eqnarray}
\d_{\beta}\d_{\alpha}\X =&&\Gamma^{\rho}_{\beta\alpha}\d_{\rho}\X
+ K_{\beta\alpha}\nn \\
\d_{\alpha}\nn =&& - g^{\beta\rho}K_{\alpha\beta}\d_{\rho}\X. 
\end{eqnarray}

The first of these equations may be taken as the the definition of the
extrinsic curvature $K$, or second fundamental form of the immersion,
while the second follows from consistency with the relations
$\nn\cdot\nn=1$ and $\d_{\alpha}\X\cdot\nn=0$. Notice that multiplying
the first of this equations by $\d_{\gamma}\X$ one readily obtains
that the connection coefficients $\Gamma$ are the ones of the
Levi-Civita connection associated with the induced metric;
multiplication by $\nn$ implies that $K$ is a symmetric tensor.

The Codazzi-Mainardi equation is obtained from
\begin{equation}
\d_{\gamma}\X\cdot (\epsilon^{\alpha\beta}\d_{\alpha}\d_{\beta}\nn )=0,
\end{equation}
which yields that $\nabla_{[\alpha}K_{\beta ]\gamma} =0$.
And finally the Gauss equation is obtained from
\begin{equation}
\d_{\gamma}\X \cdot
(\epsilon^{\rho\beta}\d_{\rho}\d_{\beta}\d_{\alpha}\X)=0,
\end{equation}
which implies that $R_{\gamma\alpha\rho\beta}= K_{\gamma [\rho}
K_{\beta ]\alpha}$, where $R$ is the Riemann curvature tensor
associated with the induced metric.

It is now intuitively clear that given two symmetric tensors $g$ and
$K$ obeying the integrability condition one may recover, up to
Euclidean motions\footnote{This is due to the fact that the first and
second fundamental forms, as defined above, are invariant under global
translations and rotations in $\R^3$.}, the associated surface by
integrating the structural equations.

One may define now the mean curvature, $H$, and the Gaussian
curvature, $K$, by
\begin{equation}
H = {1\over 2}g^{\alpha\beta} K_{\alpha\beta},\quad{\rm and}\quad
K = {1\over 2} \epsilon^{\alpha\rho}\epsilon^{\beta\gamma}
K_{\alpha\beta}K_{\rho\gamma}.
\end{equation}

\subsection{The Enneper-Weierstrass representation of minimal surfaces}

Minimal surfaces are defined via the condition $H=0$. They owe their name
to the fact that they minimize the area functional, and therefore are 
solutions to the Nambu-Goto variational problem.

It will be useful in the following to introduce an isothermal coordinate
system.  Isothermal coordinates are defined trough the condition that
the induced metric is proportional to the standard flat two-dimensional
Euclidean metric. That such a coordinate system can always be locally
achieved is a standard result which proof can be found, for example,
in \cite{Spivak}. If we denote such a coordinate system by $(x,y)$ the
zero mean curvature condition reads
\begin{equation}
{\d^2\X\over \d x^2} +{\d^2\X\over \d y^2} =0, 
\end{equation}
which we can also write as
\begin{equation}
{\d\ \over\d x}\left( {\d\X\over \d x}\right) =
{\d\ \over\d y}\left( -{\d\X\over \d y}\right). 
\end{equation}
Therefore there is locally a vectorial function $\Y$ such that
\begin{equation}
{\d \Y\over \d x} = - {\d\X\over \d y} \quad {\rm and} \quad
{\d \Y\over \d y} =  {\d\X\over \d x}. 
\end{equation}
But these are none other than the Cauchy-Riemann equations for
$\Z =\X + i\Y$.  Therefore the minimality condition implies
that $\X$ is the real part of a complex analytic function
$\Z$. Notice, however, that the condition that $(x,y)$
constitute an isothermal coordinate function yields a 
further constraint in $\Z$, i.e.,
\begin{equation}
{\d\Z\over \d z}\cdot{\d\Z\over \d z}=0,
\end{equation}
where $z =x+ i y$. Expanding in components the above expression one directly
gets the standard conditions
\begin{equation}
{\d\X\over \d x}\cdot{\d\X\over \d x}=
{\d\X\over \d y}\cdot{\d\X\over \d y} \quad {\rm and} \quad
{\d\X\over \d x}\cdot{\d\X\over \d y}=0.
\end{equation}

It is therefore natural to introduce the complex analytic function
$\PP= \partial \Z$, where $\d$ stands as a shorthand for $\d /\d z$.
Then one can naturally associate to every minimal surface
a quadric in $\C^3$ defined trough $\PP\cdot\PP =0$, and
viceversa via
\begin{equation}
\X (z,\bar z )= {\rm Re}\int^z \PP (\omega ) d\omega , 
\end{equation}
up to an arbitrary translation.

The Enneper-Weierstrass representation is now achieved by finding
an explicit solution for the quadratic equation in $\PP$. It is
straightforward to check that
\begin{equation}
\psi_1 = {1\over 2} f (1 - g^2), \quad
\psi_2 = {i\over 2} f (1 + g^2), \quad 
\psi_3 = fg,
\end{equation}
with $g$ a meromorphic function and $f$ complex analytic, and such
that it has a zero of order $2n$ wherever $g$ has a pole of order $n$,
is a solution of $\PP\cdot\PP =0$. The converse is equally simple to
prove. Notice that the quadratic equation can be written as
\begin{equation}
(\psi_1 - i\psi_2)(\psi_1 + i\psi_2 ) = -\psi_3^2.
\end{equation}
If $\psi_3$ is the zero function then we choose
$g=0$ and $f =2\psi_1$. If $\psi_3$ is not the zero function then
$\psi_1 - i\psi_2$ is also not the zero function and then one can
define
\begin{equation}
f = \psi_1 -i\psi_2, \quad g={\psi_3 \over \psi_1 -i\psi_2 },
\end{equation}
with $f$ analytic and $g$ meromorphic. Moreover it follows that
\begin{equation}
\psi_1 + i\psi_2 =-{\psi_3^2 \over \psi_1 -i\psi_2 }=-fg^2.
\end{equation}
Thus the analyticity of left hand side of the above equation implies the
condition on the zeros and poles of $f$ and $g$ to be the one stated
above. So we finally arrive to the Enneper-Weierstrass representation of
minimal surfaces. Explicitly
\begin{eqnarray}
X^1(z,\bar z)=&&{\rm Re}\int^z {1\over 2} f(\omega )(1 - g(\omega)^2)+x^1,\\
X^2(z,\bar z)=&&{\rm Re}\int^z {i\over 2} f(\omega )(1 + g(\omega)^2)+x^2,\\
X^3(z,\bar z)=&&{\rm Re}\int^z f(\omega) g(\omega )+ x^3,
\end{eqnarray}
with the $x^{i}$ real constants.

One can do still better, however, as I will now go on to show. Before
doing so one should introduce a further geometrical construction: the
Gauss map.  The Gauss map is given by associating to each point in the
surface its unit normal vector $\nn$.  This, of course, requires the
choice of an orientation, therefore we will restrict from now on our
considerations to orientable surfaces. A direct computation yields
\begin{equation}
\nn = \left( {2 {\rm Re}\, g(z)\over |g(z)|^2 + 1},
{2 {\rm Im}\, g(z)\over |g(z)|^2 + 1},
{|g(z)|^2 - 1\over |g(z)|^2 + 1}\right)\in S^2.
\end{equation}
Notice that as we approach a pole of $g(z)$ $\nn\rightarrow (0,0,1)$.

If we are not at a flat point of the surface it is possible to
fix the conformal structure by choosing as our local coordinate
the image under stereographic projection of the two-sphere onto
the complex plane plus the point at infinity. This is equivalent
to fixing $g(\omega )=\omega$. Therefore one may write
\begin{equation}
\X (z,\bar z ) = {1\over 2} {\rm Re}\int^z f(\omega )
\left( 1 - \omega^2, i (1 +\omega^2), 2\omega\right) +\x.
\end{equation}

One may now compute all the geometrically relevant quantities
in terms of $f$. In particular
\begin{equation}
g_{z\bar z}= {1\over 2} f\bar f (1 + z\bar z )^2.
\end{equation}
More interesting for our interest will turn out to be the Hopf
quadratic differential, also called skew curvature, which is
nothing but the $zz$ component of the extrinsic curvature.  A direct
computation yields
\begin{equation}
K_{zz} = \partial\partial \X\cdot \nn = f,
\end{equation}
thus giving a direct geometrical interpretation to the analytic function
$f$. As we will now see this result will show to be of the utmost
importance. Notice also that the analyticity of the skew curvature is
a direct consequence of the Codazzi-Mainardi equation restricted to
minimal surfaces.

It will be of crucial importance for the string physics to understand
how Euclidean invariance is realized in the Enneper-Weierstrass
parameterization of minimal surfaces. Notice that this
parameterization requires a soldering between target space degrees of
freedom and world-sheet ones: we are using the stereographic
projection of the unit normal vector to the surface to fix the
conformal coordinates on the surface. It is therefore natural to
expect that rotations in the target space are naturally associated
with rotation of the Riemann sphere. Indeed from the formula above it
may appear that $f$ is invariant under rotations because all indices
are properly contracted, nevertheless in order to have the surface
parameterized through the Gauss map one should take a compensating
transformation in the $z$ coordinate. It is a simple exercise to check
that rotations of the Riemann sphere correspond to an $SU(2)$ subgroup
of $SL (2,\C )$ given by
\begin{equation}
\tilde\eta= {a\eta -b\over \bar b\eta +\bar a}\quad{\rm with}
\quad a\bar a +b\bar b =1.
\end{equation}
Because $f$ is a quadratic differential it should transform such that
\begin{equation}
\tilde f(\tilde\eta ) d^2\tilde\eta = f(\eta )d^2\eta,
\end{equation}
or explicitly for $SU(2)$ transformations
\begin{equation}
\tilde f(\tilde\eta (\eta )) = (\bar b\eta + \bar a)^4 f(\eta ).
\end{equation}
Let me check that the transformed $f$ does indeed correspond to a
rotated surface. For the time being I will ignore the zero modes that
transform in the usual way. If one considers
\begin{equation}
\tilde\X (z,\bar z ) = {1\over 2}{\rm Re} \int^z d\tilde\eta 
\tilde f (\tilde\eta)
\left( \begin{array}{c}
1-\tilde\eta^2 \\ 
i(1 +\tilde\eta^2)\\ 
2\tilde\eta 
\end{array} \right)
\end{equation}
one may change variables in the integral to obtain
\begin{equation}
\tilde\X (z, \bar z)= {1\over 2}{\rm Re} \int^{\tilde z}
d\eta f(\eta)
\left( \begin{array}{c}
(\bar b\eta +\bar a)^2 - (a\eta -b)^2\\ 
i(\bar b\eta +\bar a)^2 +i (a\eta -b)^2\\ 
2(\bar b\eta +\bar a)^2 (a\eta -b)^2\\ 
\end{array} \right),
\end{equation}
with
\begin{equation}
\tilde z = {\bar a z +b\over -\bar b z +a}.
\end{equation}
If one chooses $a={\rm e}^{i\theta/2}$ and $b=0$, after a little
algebra one arrives to the expression
\begin{equation}
\tilde\X = {1\over 2}{\rm Re} \int^{\tilde z}
d\eta f(\eta)
\left( \begin{array}{c}
\cos\theta (1-\eta^2) - i\sin \theta (1 +\eta^2)\\ 
\sin\theta (1-\eta^2) + i\cos \theta (1 +\eta^2)\\ 
2\eta\\ 
\end{array} \right).
\end{equation}
Therefore one finally obtains
\begin{equation}
\tilde\X (z, \bar z) = {\mathcal R}(\ee_3,\theta )\X (\tilde
z,\tilde{\bar z}),
\end{equation}
with ${\mathcal R}(\ee_3, \theta )$ being the standard $SO(3)$ matrix
associated with a rotation of an angle $\theta$ around the axis
given by $\ee_3$.

A little more of work shows that the case with $a=\cos\phi/2$ and
$b=\sin\phi/2$ corresponds to a rotation of angle $\phi$ around
$\ee_2$; but as it is well-known, these two rotations generate the
whole $SO(3)$ group of rotations, therefore showing the correctness
of our assumption.

\section{The Nambu-Goto action}

Although this is by no means a review in string theory, the
purpose of this section is
to recall certain properties of the Nambu-Goto
string that will be extensively used in the following, as well as to set up
my notations and conventions.

The Nambu-Goto action is the simplest geometric invariant of an
immersed surface, i.e., its area. 
\begin{equation}
S(\Sigma )= -{1\over 4\pi\alpha}{\rm Area}(\Sigma )
\end{equation}
As already mentioned in the previous section, the solution of its
associated variational problem is given by surfaces of zero mean
curvature. In the conformal gauge, or equivalently isothermal
coordinates, the solution takes the general form (I will work in a
system of units in which $\alpha = 1/2$)
\begin{equation}
\X (z,\bar z)=
\x - {i\over 4}\p {\rm ln}\, z\bar z + {i\over 2}\sum_{n\neq 0}{1\over n}
\balpha_n z^{-n} 
+{i\over 2}\sum_{n\neq 0}{1\over n}
\bbalpha_n \bar z^{-n}
\end{equation}
if one chooses periodic boundary conditions.  But, as it is well
known, extra constraints come from the reparameterization invariance
of the action. They can be written in terms of $\X$ as
\begin{equation}
\partial\X\cdot\partial\X =\bar\partial\X\cdot\bar\partial\X =0.
\end{equation}
I will restrict myself, from now on, to the closed string case. The
geometrical reasons to do so is to avoid flat surfaces as
solutions. Notice that the geometrical parameterization of the surface
by its Gauss map requires to rule out that case, which is only a
priori allowed for open string boundary conditions.  With all of this
in mind, one can now directly apply all the machinery developed in the
previous chapter.

It will be convenient to split the field $\X$ into its
holomorphic and antiholomorphic parts as $\X(z,\bar z) =
\X (z) + \bar\X (\bar z)$ with
\begin{eqnarray}
\X (z) =&& {1\over 2}\x - {i\over 4}\p {\rm ln}\, z 
+ {i\over 2}\sum_{n\neq 0}{1\over n}\balpha_n z^{-n} ,\\
\bar\X (\bar z) =&& {1\over 2}\x - {i\over 4}\p {\rm ln}\, \bar z 
+ {i\over 2}\sum_{n\neq 0}{1\over n}\bbalpha_n \bar z^{-n} .
\end{eqnarray}

In terms of these fields one may write the Enneper-Weierstrass map as
\begin{equation}
\X (z) = {1\over 2} \int^z f(\omega )
\left( 1 - \omega^2, i (1 +\omega^2), 2\omega\right) +
{1\over 2}\x,
\end{equation}
and the obvious equivalent expression for $\bar\X (\bar z )$.
{}From now on I will concentrate almost exclusively in the holomorphic
part, the results being trivially extended to the antiholomorphic
sector. 
\section{The covariant phase space formalism}

The original idea of developing the canonical formalism in an
explicitly covariant manner is first due to Witten \cite{Witten}, and
later developed by Crnkovic \cite{Crnkovic} and Zuckerman
\cite{Zuckerman}.  The idea is simple and is based on the observation
that there is a one-to-one relationship among points in the phase
space and solutions to the equation of motion. This is roughly
equivalent to say that given initial conditions, that require a
noncovariant choice of a space-time slice, the solution to the
equations of motion is fully determined (of course, special care
should be taken in the presence of gauge invariances). It is therefore
natural in a covariant field theory to preserve explicitly the
covariance properties of the theory and define the phase space
directly as the solution space of the associated field theory.

Our final goal is to define Poisson brackets in the reduced phase space
associated with the three-dimensional Nambu-Goto string. In order to do
so I will have to define some basic objects: functions, vector fields
and differential forms living in an infinite-dimensional phase space.
To avoid getting into analytical details it will be useful to algebraize
the necessary notions. This point of view may not be familiar to everyone,
so it may be useful to introduce it firstly in the familiar setting of 
(finite-dimensional)
classical hamiltonian dynamics.

Let me start with a simple example. Consider $M = \R^{2n}$ for
our phase space with coordinates $(q^{i},p_i)$. The Poisson bracket
of any two functions $f$ and $g$ is given by
\begin{equation}
\{ f,g \} =\sum_i \left( {\d f\over \d q^{i}}{\d g\over \d p_i} -
{\d f\over \d p_i} {\d g\over \d q^{i}}\right).
\end{equation}
It is easy to see that the Poisson bracket is antisymmetric and that it
satisfies the Jacobi identity. Therefore it gives the ring of functions
on $M$ the structure of a Lie algebra.  More is true, however. The Poisson
bracket is also easily seen to act as a derivation on the ring of functions:
if $f$, $g$, and $h$ are functions on $M$,
\begin{equation}
\{ f, gh \} = g \{ f,h\} + \{f,g\} h.
\end{equation}
These facts turn the functions on $M$ into a Poisson algebra.

Suppose that we now change coordinates to $x^{i} (q,p)$. The fundamental
Poisson bracket of these coordinates is given by
\begin{equation}
\Omega^{ij} = \{x^{i},x^j\} =
\sum_k \left( {\d x^{i}\over \d q^{k}}{\d x^j\over \d p_k} -
{\d x^{i}\over \d p_k} {\d x^j\over \d q^{k}}\right).
\end{equation}
It is easy to check that $\Omega^{ij}$ transforms tensorially under an
arbitrary change of coordinates and thus defines an antisymmetric
bivector; That is, a rank 2 antisymmetric covariant tensor. Furthermore,
one can check that $\Omega^{ij}$ is nondegenerate so that its inverse
$\Omega_{ij}$ exists and defines a nondegenerate 2-form on $M$, called
the symplectic form. The Jacobi identities of the Poisson bracket imply
a differential relation on $\Omega^{ij}$ which, when inverted, imply
that the symplectic form is closed.

To summarize, starting with the usual coordinates $(q,p)$ and the usual
Poisson brackets, we have uncovered an underlying geometric structure:
a non-degenerate closed 2-form. This may seem overkill for $\R^{2n}$
but it allows us to define a Poisson structure on any manifold $M$
possessing a symplectic form. Of course, around each point of $M$, we can
choose coordinates $(q,p)$ whose Poisson bracket are the standard ones.
This is the essence of Darboux's theorem. It states that symplectic manifolds
of the same dimension are locally isomorphic.

The Poisson bracket allows us to assign to every function $f$ a 
hamiltonian vector
field $H_f$ as follows:
\begin{equation}
H_f\cdot g =\{ f, g\},
\end{equation}
whose components in local coordinates are given by $H^{i}_f =
-\Omega^{ij}\partial_j f$. Therefore the Poisson brackets can be 
directly written in terms of the symplectic form $\Omega$
as
\begin{equation}
\{ f ,g \} = \Omega (H_f ,H_g),
\end{equation}
with
\begin{equation}
\Omega (H_f,  \,\cdot\, ) = - df \,\cdot,
\end{equation}
as a direct computation in component reveals.

Beucase the symplectic form is a closed 2-form it is always possible,
at least locally, to define a 1-form $\theta$, usually called the
canonical 1-form, such that $\Omega =d\theta$. It is then possible to
define the Poisson brackets directly in terms of the canonical 1-form
with the help of the Lie formula, i.e.,
\begin{equation}
\{ f ,g \} = d\theta (H_f ,H_g) = H_f\cdot\theta (H_g) -
H_g\cdot\theta (H_f) - \theta ([ H_f,H_g ]).
\end{equation}

We will still need another piece of geometrical information regarding
presymplectic rather than symplectic manifolds. It is usual in dynamical
systems to work with coordinates in phase space that are subject to 
constraints. Although physicists are well acquainted with Dirac's
treatment of constraints in Lagrangian systems, for the case at hand
we will need a more geometrical, though equivalent, 
approach to the subject. Fortunately,
the ideas and techniques involved are still simple enough to be presented 
succinctly. 

I will concentrate in the reduction to a submanifold of the
original phase space defined by second class constraints; a symplectic
submanifold in the standard geometrical nomenclature \cite{Woodhouse}.
In this particular case the reduction process is trivial: if we denote
by $C$ the symplectic submanifold, one can naturally define
a symplectic form on it simply by restriction of the 
original 2-form $\Omega$,
which is usually denoted by $\Omega |_C$. More precisely, if $Z$ and
$Y$ belong to the tangent space at a point $p$ of $C$ 
\begin{equation}
\Omega|_C (Z,Y) = \Omega (Z,Y),
\end{equation}
where in the right hand side $Z$ and $Y$ are considered as elements of
$T_pM$.
It is somehow more tedious,
although straightforward,
to show that this natural geometric construction corresponds to the
standard Dirac bracket prescription. The interested reader can
find a proof of this fact in \cite{Diracs}.

It will be more convenient
to think of the reduced phase space in a more intrinsic way and
look at $C$ as the embedding of the physical phase space $M_0$ into
$M$. If we denote the embedding map by $\varphi$ one can define
a symplectic 2-form $\omega$ in $M_0$ acting on any two vector
fields $Z,Y\in TM_0$ by
\begin{equation}
\omega (Z,Y)=\varphi^{*}\Omega (Z,Y)=\Omega (\varphi_{*}Z,
\varphi_{*}Y),
\end{equation}
where $\varphi_{*}Z$ stands for the pushforward of the vector field
$Z$ in $M_0$ trough the map $\varphi$.

With all of this in mind we can now go on to analyze the case at hand.
Rather than describing the general formalism that allows the
construction of symplectic structures in the solution space of a
dynamical system, I will work in detail the example in which we are
interested at present.

The solution space  associated with
the Nambu-Goto equation of motions can be described as the solution
space of the equation
\begin{equation}
\partial\bar\partial\X =0,
\end{equation}
with adequate boundary conditions, more on this will follow, and
subject to the constraints
\begin{equation}
\d \X\cdot\d \X = \bar\d \X\cdot\bar\d \X =0.
\end{equation}

In order to construct the symplectic structure I will first
concentrate in the unreduced manifold $M$, i.e., solutions of the
equations $\bar\d\d\X$.  Like in the finite-dimensional case one
should first identify the functions, vector fields and one-forms in
$M$. The functions, or rather functionals, are simply maps from the
space $M$ into the reals. A canonical example is supplied by the
evaluation map, that is obtained by choosing a space-time point and
evaluating a solution of the equation of motion in that point. We will
be particularly interested in linear functionals.  An example of them
is supplied by the weighted integral in $\R^3$ of a solution to the
equation of motion, i.e.,
\begin{equation}
F_{\N }(\X ) =\int_{\Sigma}\N\cdot\X,
\end{equation}
although other natural examples will pop up in the following.

Vector fields are naturally parameterized by deformations of the solutions
preserving the equations of motion, i.e., symmetries. In our case,
since the equations defining $M$ are linear, vector fields are
themselves parameterized by solutions of the equation itself.  Their
action on functions is defined as usual by
\begin{equation}
\partial_{\A} G(\X )= {d\ \over d\epsilon} 
G(\X + \epsilon\A )|_{\epsilon =0},
\end{equation}
where $\bar\d\d\A =0$.

Finally we should define the gradient of a function $G$ or its associated
1-form $dG$. One can easily do so by defining its action on vector
fields to be
\begin{equation}
dG (\partial_{\A})=\partial_{\A} G,
\end{equation}
mimicking the standard finite-dimensional definition.  From now on,
and in order not to clutter up the notation, I will confuse the vector
fields with the solutions of the equation of motion parameterizing
them.  It will be convenient to center our attention in the
holomorphic sector. It will be useful to explicitly split the zero
modes and write an holomorphic deformation $\A(z)$ as
\begin{equation}
\A (z) = {1\over 2}\asf -{i\over 4}\A_0 {\rm ln}z + \A_z (z),
\end{equation}
with
\begin{equation}
\A_z = {i\over 2} \sum_{n\neq 0} {1\over n}\A_n z^{-n}.
\end{equation}

Let me show now that the following canonical 1-form $\theta$ in $M$
\begin{equation}
\theta ({\A} )={1\over \pi}\oint dz\,\partial\X\cdot\A_z
+ \, \p\cdot\asf, 
\end{equation}
does the required job,
where $\oint$ stands for the Cauchy integral for a contour that has $z=0$
as an interior point, and we are considering closed string boundary
conditions for $\X$ and consequently for $\A$. One should now compute,
with the use of the Lie formula, the explicit expression for the
symplectic form $\omega$, but rather than writing the general
expression I will restrict myself to the case of constant vector
fields in the covariant phase space, i.e., vector fields that are
field independent. For that particular case a direct computation
yields
\begin{equation}
\Omega (\A ,\B) ={2\over \pi}\oint dz\,\partial\A_z\cdot\B_z
+( \A_0\cdot\bsf - \asf\cdot\B_0 ). 
\end{equation}

It is now simple to check that the 2-form defined above
induces the usual Poisson brackets among the modes of the string.
If one considers the linear functions
\begin{equation}
F_p^i = {1\over\pi}\oint dz\, z^p \partial X^i,
\end{equation}
one directly obtains that $F_p^i = \alpha_p^i$ for
$p\neq 0$. Therefore
\begin{equation}
\{ F_p^i, F_r^j\} =\{\alpha_p^i, \alpha_r^j \}.
\end{equation}
In order to compute this Poisson brackets one should first compute
the hamiltonian vector fields associated with these linear functionals.
For example, $\HH_{F_p^i}$ 
it is obtained from
\begin{equation}
\Omega ( \HH_{F_p^i}, \B ) = -d F_p^i(\B) =-
{1\over\pi}\oint z^p \partial B^i,
\end{equation}
which, together with the definition of $\Omega$, automatically implies that
\begin{equation}
\HH_{F_p^i} = {1\over 2} z^p \ee_i,
\end{equation}
where the $\ee_i$'s form the standard basis for vectors in $\R^3$. From
here directly follows that
\begin{equation}
\{\alpha_p^i, \alpha_r^j \}= dF_r^j (\HH_{F_p^i})={1\over 2\pi }\oint
z^r \partial z^p\, \delta_{ij} = i p\,\delta^{ij}\delta_{p+r, 0}.
\end{equation}
A similar computation yields the lacking Poisson bracket
\begin{equation}
\{ p^i, x^j \} =\delta^{ij},
\end{equation}
with all the remaining Poisson brackets being zero.
Notice that under analytic continuation to Minkowski space,
with our conventions, $p^3$ and $x^3$ pick up an extra factor of
$i$ and therefore one obtains the standard Minkowskian Poisson
brackets
$\{ p^i, x^j \} =\eta^{ij}$. 
Thus proving the equivalence of the covariant phase space approach and
standard symplectic methods for the case at hand.

\subsection{The Poisson brackets in the reduced phase space}

Of course, this long detour has not been a caprice and I will go on
now to show how the previous developments will allow us to compute the
Poisson brackets in the reduced phase space. We are interested in
computing the fundamental induced Poisson brackets on the modes of
$f(z)$. If we define
\begin{equation}
f(z) = \sum_{n\in\Z} {f_n\over z^{n +2}},
\end{equation}
we will be naturally interested in the following functions
\begin{equation}
G_p ={1\over 2\pi i}\oint dz z^{p+1} f(z).
\end{equation}
Let me first consider the case for which $p\neq 0,\pm 1$.
As before one should start by finding the explicit expression of the
hamiltonian vector field associated with $G_p$. From the definitions
one directly gets
\begin{equation}
\omega (H_{G_p}, \xi ) = \varphi^*\Omega (H_{G_p}, \xi)=
-dG_p (\xi ) = -{1\over 2\pi i}\oint dz z^{p+1}\xi (z).
\end{equation}	
The definition of the Enneper-Weierstrass map together with the
2-form $\Omega$ defined above yield
\begin{equation}
\varphi^*\Omega (H_{G_p}, \xi ) = {1\over \pi}
\oint dz \xi (z) \int^z d\eta H_{G_p}(\eta ) (z -\eta )^2.
\end{equation}
{}From where one obtains that
\begin{equation}
H_{G_p} = {i\over 4} p (p^2 -1) z^{p-2}.
\end{equation}
Therefore from this follows that
\begin{equation}
\{G_p, G_q\}=\{f_p , f_q\} = H_{G_p}\cdot G_q ={p (p^2 -1)\over 8\pi}
\oint dz z^{q+1}z^{p-2}, 
\end{equation}
or
\begin{equation}
\{ f_p, f_q\} = {i\over 4}p (p^2 -1)\delta_{p+q,0}
\end{equation}
for $p\neq 0,\pm 1$.

In order to compute the remaining Poisson brackets it will show
convenient to introduce the following notation: 
\begin{equation}
p^1 = 2 i(f_1 -f_{-1}),\quad p^2 = -2  (f_1 +f_{-1}),\ {\rm and}\
p^3 = 4i f_0.\label{eq:momentum}
\end{equation}
It is now an straightforward computation to show that
\begin{equation}
\{ p^{i}, x^{j}\} =\delta^{ij},
\end{equation}
with all other remaining Poisson brackets being zero.
Notice that for the momentum associated with the zero modes the 
Enneper-Weierstrass is one to one, the above result
directly reflects that fact, as can be directly checked from the
definition of $\X$ in terms of the $f_j$.

\section{The quantum theory}

Now we have all the required information to construct the associated
quantum theory. The first step is trivial and correspond to the
standard rule of substituting Poisson brackets for commutators,
i.e., $[\ ,\ ] =-i \{\ ,\ \}$. 
In our case the Heisenberg quantization rule ($\hbar =1$) yields
\begin{equation}
[ \hat f_p ,\hat f_q ] = {1\over 4} p (p^2 -1) \delta_{p+q,0},
\end{equation}
and an identical expression for the modes of $\hat{\bar f_n}$,
together with
\begin{equation}
[ \hat x^{i},\hat p^{j} ]= i\delta^{ij}.
\end{equation}
Notice that the mode algebra of the $\hat f_n$ corresponds to the central
term of the Virasoro algebra for $D=3$. Although, at this point, I do
not know if this is something more than sheer coincidence.

The `in' vacuum state $|0 \rangle$ is defined trough the conditions
\begin{equation}
\hat f_n |0\rangle =0\quad {\rm for}\quad n\geq -1.
\end{equation}
All states of momentum $p$ are obtained from the action of creation
operators, i.e., $\hat f_n$ with $n\leq -2$, acting on the state
$|0,p\rangle$ which is defined to be annihilated by all $\hat f_n$ with
$n\geq 2$ and such that
\begin{equation}
\hat p^{i} |0,p\rangle = p^{i}|0,p\rangle.
\end{equation}

The `out' vacuum state $\langle 0|$ is equally easily defined by
taking the adjoint of $\hat f_n$ to be
\begin{equation}
\hat f^\dagger_n =\hat f_{-n},
\end{equation}
which corresponds with
\begin{equation}
\hat f^{\dagger} (z) = {1\over \bar z^4}{\hat f({1\over\bar z})}.
\end{equation}

The reader may worry about the fact that although this rule seems
natural from the conformal field theory point of view it may be
unwarranted in this case, where the conformal invariance has been
explicitly broken.  But once again the geometry of surfaces comes to
our rescue.  In fact it is a simple exercise to check that the surface
defined trough an $\bar f(\bar z)$ given by $f^{\dagger}$, as defined
above, is the same as the one obtained from the original $f$ up to a
change in parameterization given by $z\rightarrow 1/\bar z$ plus a
reflection in the $X^3$ coordinate.  But, let me recall, that from the
Euclidean field theory point of view the adjoint operation requires a
time inversion due to the lack of a factor of $i$ in the evolution
operator. In radial quantization this is equivalent to the
transformation $z\rightarrow 1/\bar z$ \cite{Ginsparg}. The extra sign
factor in the third coordinate can be equally understood in terms of
the extra factor of $i$ required to pass from Euclidean to Minkowskian
signature in the target space-time.  One may equally check that this
hermitian conjugation rule comes naturally imposed form the usual one
of the string modes in Minkowski space-time.

Notice also that this definition of adjoint imply
that the momentum operators $p^1, p^2$, and
$ip^3$, as defined from
equation (\ref{eq:momentum}), are selfadjoint.

More importantly the above adjoint properties of the operator
$\hat f$ imply that the inner product of its associated
Hilbert spaces, i.e., the ones obtained by the repeated
application of creation operators onto the reference state
$|0,p\rangle$, is positive-definite. A crucial property of the quantum
theory required for consistency with the standard probabilistic
interpretation of the transition amplitudes.
 
\subsection{Quantum realization of Euclidean invariance}

We have already studied in previous sections how rotational 
invariance was implemented at the classical level
within the Enneper-Weierstrass representation of minimal surfaces.
In our case, and in contrast with other gauge fixings like
the light-cone one, the implementation of the Euclidean invariance
\footnote{Translational symmetry is simply implemented by shifts
in the zero modes.} at
the quantum level is completely straightforward.
In fact it is a simple computation to show that the following
operators
\begin{eqnarray}
J^3=&& -4 \sum_{n\geq 2} {1\over n^2 -1} : \hat f_{-n}
\hat f_{n}: \\
J^2 =&& 2i \sum_{n\geq 2} {1\over n(n+1)} : \hat f_{-n-1} \hat f_{n}: 
 - 2i\sum_{n\geq 2} {1\over n(n-1)} : \hat f_{-n+1} \hat f_{n}:\\ 
J^1 =&& -2 \sum_{n\geq 2} {1\over n(n+1)} : \hat f_{-n-1} \hat f_{n}: 
 - 2\sum_{n\geq 2} {1\over n(n-1)} : \hat f_{-n+1} \hat f_{n}:
\end{eqnarray}
obey the standard $SU(2)$ algebra $ [J^{i}, J^j]=i
\epsilon^{ijk}J^k$, and moreover generate the right transformations
on the modes of $\hat f$. Of course, the $x^j$ and $p^j$ operators
transform in the standard fashion, and I will omit the complete
expression of the $J$'s involving them and 
the modes of the operator associated
with $\bar f$.

It is also instructive to notice that Euclidean invariance of the two
point correlation function for the $\hat f$ operator requires that
\begin{equation}
\langle \hat f(z) \hat f(0)\rangle = {c/2\over z^4}.
\end{equation}
That is, of course, the result obtained in this case with
$c=3$.

All of this completes the proof that the Nambu-Goto string
when quantized in the reduced covariant phase space approach
yields a consistent and Euclidean-invariant, two-dimensional
quantum field theory.

\subsection{Particle states and spectrum}

The physical states are obtained by the repeated action of the creation
modes of the fields $f$ and $\bar f$ on the momentum state 
$| 0, p\rangle$. They will be labeled by two multi-indices as follows
\begin{equation}
|\{ j,\bar j\},p\rangle = \hat f_{-j_1}\cdots \hat f_{-j_n}
\hat{ \bar{f}}_{-\bar{j}_1}\cdots \hat{\bar{f}}_{-\bar{j}_n} |0,p\rangle
\end{equation}

It is simple to show now that they carry irreducible representations
of the Euclidean group associated with particles of integer spin and
arbitrary mass. In order to do so, it will be convenient to recall
some basic facts about Wigner method of induced representations
applied to the three-dimensional Euclidean group.

One should start by identifying the Casimir operators. They are given
by $p^2$ and $S=p\cdot J$. I will denote by $m^2$ and $sm$ its
respective eigenvalues, with $m$ the mass of the particle state and
$s$ its spin. Therefore our particle states will be labeled by its
mass and spin.

The components of the momentum $p^j$ commute among themselves, so it
is natural to label the physical states in terms of their
eigenvectors. It is now standard to obtain the irreducible
representations of the Euclidean group by choosing a standard momentum
$k^j$ and inducing the representations of the full Euclidean group
from those of the stability subgroup associated to that particular
momentum $k^j$, i.e., its little group. Let me reproduce the standard
procedure for the case when $k=(0,0,m)$, although, of course, any
other momentum will do.  The little group leaving invariant this
reference momentum is one-dimensional and its algebra is generated by
$J^3$. Its irreducible representations are given by a phase ${\rm
e}^{-is\omega}$, where $\omega$ is the angle of rotation. The action
of any rotation $\Lambda$ on any state $| p,s\rangle$ can be obtained
in an standard fashion as follows.  Let me denote by $R$ the rotation
such that
\begin{equation}
p^j = R^j_i (p) k^{i}.
\end{equation}
We can then define the state $|p,s\rangle$ by
\begin{equation}
|p,s\rangle \sim U(R(p))|k,s\rangle,
\end{equation}
up to an unimportant normalization factor.

The action of an arbitrary rotation $\Lambda$
may be now easily computed as follows.
\begin{equation}
U(\Lambda )|p,s\rangle \sim U(\Lambda R(p)) |k,s\rangle 
\sim U(R(\Lambda p)) U( R^{-1}(\Lambda p) \Lambda R(p)) |k,s\rangle. 
\end{equation} 
The rationale behind this last step being that $ U( R^{-1}(\Lambda p)
\Lambda R(p))$ leaves invariant the momentum $k$,
i.e., it belongs to its little group. As commented above its
irreducible representations are one-dimensional and one obtains
\begin{equation}
U(\Lambda )|p,s\rangle \sim {\rm e}^{-is \omega_{\Lambda}}
 |\Lambda p, s\rangle ,
\end{equation}
where $\omega_{\Lambda,p}$ is the Wigner rotation angle, and can
be directly computed from the above expressions.

{}From all of this and the definition of $J^3$ it now follows that the
state $|\{ j,\bar j\},p\rangle $ corresponds to a irreducible
representation of the Euclidean group of mass $p^2$ and spin
\begin{equation}
s = \sum_{ \{ j\} } j_i - \sum_{ \{ \bar j\} } \bar{j}_i. 
\end{equation}

Notice that due to the fact that the constraints have been solved
there is no, at least to the best of my understanding, quantization
condition on the mass spectrum in this scheme. Therefore the particle
spectrum consists of particles of integer spin and arbitrary mass.

As a final comment, notice that in the quantization process I have
assumed that the Enneper-Weierstrass parameterization is globally
defined, which clearly implies that the surfaces is of genus zero,
i.e., the surfaces result from a map of the Riemann sphere into
spacetime. Therefore, in physicists' language, I have restricted
myself to the free case. There is already a vast literature on how to
extend the operator formalism to attack the interacting, or higher
genus, theory.  The application of those methods to this particular
case will be a problem for the (near?) future.

\begin{ack}
I would like to thank E. Alvarez, J. Borlaf, J.M. Figueroa-O'Farrill,
J. Gonzalo, J. Mas, P. Meessen, and J. Roca for useful conversations
regarding these matters.
\end{ack}

\end{document}